\documentclass[aps,prd,twocolumn,amsmath,amssymb,superscriptaddress]{revtex4-1}
\usepackage{graphicx, subfigure}
\usepackage{dcolumn}
\usepackage{bm}
\usepackage{hyperref}
\usepackage{color}
\usepackage{url}
\usepackage{breakurl}
\usepackage{multirow}

\begin{document}

\title{Measurement of the cosmic ray energy spectrum with IceTop-73}

\affiliation{III. Physikalisches Institut, RWTH Aachen University, D-52056 Aachen, Germany}
\affiliation{School of Chemistry \& Physics, University of Adelaide, Adelaide SA, 5005 Australia}
\affiliation{Dept.~of Physics and Astronomy, University of Alaska Anchorage, 3211 Providence Dr., Anchorage, AK 99508, USA}
\affiliation{CTSPS, Clark-Atlanta University, Atlanta, GA 30314, USA}
\affiliation{School of Physics and Center for Relativistic Astrophysics, Georgia Institute of Technology, Atlanta, GA 30332, USA}
\affiliation{Dept.~of Physics, Southern University, Baton Rouge, LA 70813, USA}
\affiliation{Dept.~of Physics, University of California, Berkeley, CA 94720, USA}
\affiliation{Lawrence Berkeley National Laboratory, Berkeley, CA 94720, USA}
\affiliation{Institut f\"ur Physik, Humboldt-Universit\"at zu Berlin, D-12489 Berlin, Germany}
\affiliation{Fakult\"at f\"ur Physik \& Astronomie, Ruhr-Universit\"at Bochum, D-44780 Bochum, Germany}
\affiliation{Physikalisches Institut, Universit\"at Bonn, Nussallee 12, D-53115 Bonn, Germany}
\affiliation{Universit\'e Libre de Bruxelles, Science Faculty CP230, B-1050 Brussels, Belgium}
\affiliation{Vrije Universiteit Brussel, Dienst ELEM, B-1050 Brussels, Belgium}
\affiliation{Dept.~of Physics, Chiba University, Chiba 263-8522, Japan}
\affiliation{Dept.~of Physics and Astronomy, University of Canterbury, Private Bag 4800, Christchurch, New Zealand}
\affiliation{Dept.~of Physics, University of Maryland, College Park, MD 20742, USA}
\affiliation{Dept.~of Physics and Center for Cosmology and Astro-Particle Physics, Ohio State University, Columbus, OH 43210, USA}
\affiliation{Dept.~of Astronomy, Ohio State University, Columbus, OH 43210, USA}
\affiliation{Dept.~of Physics, TU Dortmund University, D-44221 Dortmund, Germany}
\affiliation{Dept.~of Physics, University of Alberta, Edmonton, Alberta, Canada T6G 2E1}
\affiliation{D\'epartement de physique nucl\'eaire et corpusculaire, Universit\'e de Gen\`eve, CH-1211 Gen\`eve, Switzerland}
\affiliation{Dept.~of Physics and Astronomy, University of Gent, B-9000 Gent, Belgium}
\affiliation{Dept.~of Physics and Astronomy, University of California, Irvine, CA 92697, USA}
\affiliation{Laboratory for High Energy Physics, \'Ecole Polytechnique F\'ed\'erale, CH-1015 Lausanne, Switzerland}
\affiliation{Dept.~of Physics and Astronomy, University of Kansas, Lawrence, KS 66045, USA}
\affiliation{Dept.~of Astronomy, University of Wisconsin, Madison, WI 53706, USA}
\affiliation{Dept.~of Physics and Wisconsin IceCube Particle Astrophysics Center, University of Wisconsin, Madison, WI 53706, USA}
\affiliation{Institute of Physics, University of Mainz, Staudinger Weg 7, D-55099 Mainz, Germany}
\affiliation{Universit\'e de Mons, 7000 Mons, Belgium}
\affiliation{T.U. Munich, D-85748 Garching, Germany}
\affiliation{Bartol Research Institute and Department of Physics and Astronomy, University of Delaware, Newark, DE 19716, USA}
\affiliation{Dept.~of Physics, University of Oxford, 1 Keble Road, Oxford OX1 3NP, UK}
\affiliation{Dept.~of Physics, University of Wisconsin, River Falls, WI 54022, USA}
\affiliation{Oskar Klein Centre and Dept.~of Physics, Stockholm University, SE-10691 Stockholm, Sweden}
\affiliation{Department of Physics and Astronomy, Stony Brook University, Stony Brook, NY 11794-3800, USA}
\affiliation{Department of Physics, Sungkyunkwan University, Suwon 440-746, Korea}
\affiliation{Dept.~of Physics and Astronomy, University of Alabama, Tuscaloosa, AL 35487, USA}
\affiliation{Dept.~of Astronomy and Astrophysics, Pennsylvania State University, University Park, PA 16802, USA}
\affiliation{Dept.~of Physics, Pennsylvania State University, University Park, PA 16802, USA}
\affiliation{Dept.~of Physics and Astronomy, Uppsala University, Box 516, S-75120 Uppsala, Sweden}
\affiliation{Dept.~of Physics, University of Wuppertal, D-42119 Wuppertal, Germany}
\affiliation{DESY, D-15735 Zeuthen, Germany}

\author{M.~G.~Aartsen}
\affiliation{School of Chemistry \& Physics, University of Adelaide, Adelaide SA, 5005 Australia}
\author{R.~Abbasi}
\affiliation{Dept.~of Physics and Wisconsin IceCube Particle Astrophysics Center, University of Wisconsin, Madison, WI 53706, USA}
\author{Y.~Abdou}
\affiliation{Dept.~of Physics and Astronomy, University of Gent, B-9000 Gent, Belgium}
\author{M.~Ackermann}
\affiliation{DESY, D-15735 Zeuthen, Germany}
\author{J.~Adams}
\affiliation{Dept.~of Physics and Astronomy, University of Canterbury, Private Bag 4800, Christchurch, New Zealand}
\author{J.~A.~Aguilar}
\affiliation{D\'epartement de physique nucl\'eaire et corpusculaire, Universit\'e de Gen\`eve, CH-1211 Gen\`eve, Switzerland}
\author{M.~Ahlers}
\affiliation{Dept.~of Physics and Wisconsin IceCube Particle Astrophysics Center, University of Wisconsin, Madison, WI 53706, USA}
\author{D.~Altmann}
\affiliation{Institut f\"ur Physik, Humboldt-Universit\"at zu Berlin, D-12489 Berlin, Germany}
\author{J.~Auffenberg}
\affiliation{Dept.~of Physics and Wisconsin IceCube Particle Astrophysics Center, University of Wisconsin, Madison, WI 53706, USA}
\author{X.~Bai}
\thanks{Physics Department, South Dakota School of Mines and Technology, Rapid City, SD 57701, USA}
\affiliation{Bartol Research Institute and Department of Physics and Astronomy, University of Delaware, Newark, DE 19716, USA}
% ...

\collaboration{IceCube Collaboration}

\begin{abstract}
 We report on the measurement of the all-particle cosmic ray energy spectrum with the IceTop air shower array in the energy range from 1.58\,PeV to 1.26\,EeV. The IceTop air shower array is the surface component of the IceCube Neutrino Observatory at the geographical South Pole. The analysis was performed using only information from IceTop. The data used in this work were taken from June 1, 2010 to May 13, 2011. During that period the IceTop array consisted of 73 stations compared to 81 in its final configuration. The measured spectrum exhibits a clear deviation from a single power law above the knee around 4\,PeV and below 1\,EeV. We observe spectral hardening around 18\,PeV and steepening around 130\,PeV.
\end{abstract}

\pacs{1,2,3}


\maketitle

\section{\label{intro} Introduction}
High resolution measurements of the cosmic ray energy spectrum and chemical composition will improve our understanding of the acceleration and propagation of high energy cosmic rays.
For cosmic ray particles with energies above some 100\,TeV this becomes a challenge since all information is derived indirectly from measurements of extensive air showers. Recently, several experiments reported spectral features or deviations from the smooth power law of cosmic ray energy spectrum between the knee at about 4\,PeV and the ankle at about 4\,EeV \cite{IT26, IC40, gamma, kascadegrande, tunka}. In this paper we investigate the spectrum in the region from 1.58\,PeV up to 1.26\,EeV. We report on the measurement of the spectrum by the IceTop air shower array in its 73 station configuration using the shower size for energy estimation and zenith dependence of the shower attenuation for estimating the uncertainty on flux due to primary composition. In section \ref{DD} the IceTop experiment and experimental data are described and simulation data are described in section \ref{sim}.  The reader is referred to reference \cite{IT_detP} for detailed technical information on the IceTop detector. The main analysis will be described in section \ref{analysis}.  

\begin{figure}[!t]
  \centering
  \includegraphics[width=3.3in]{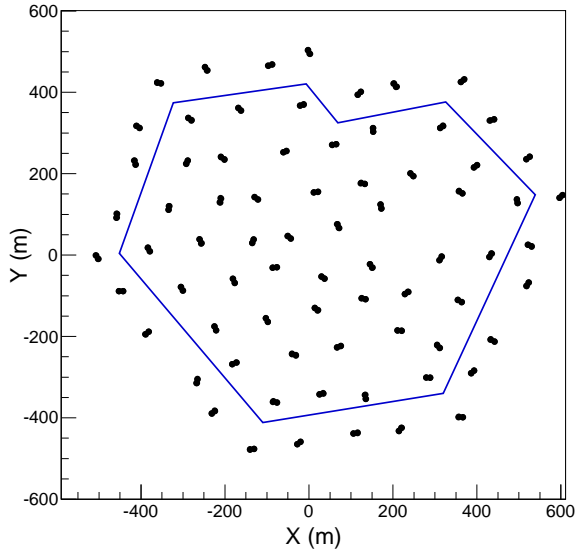}
  \caption{Surface map of IceTop in 2010. The polygon represents the containment region (577,265\,m$^{2}$).}
  \label{array}
\end{figure}

\section{\label{DD}The IceTop detector and data selection}

IceTop \cite{IT_detP} is the surface air shower array of the IceCube Neutrino Observatory at the geographical South Pole. It is located on top of the Antarctic ice sheet at an altitude of 2835\,m above sea level where the measured average atmospheric depth is 692\,g/cm$^2$. IceTop is designed to detect air showers from primary cosmic rays in the 300\,TeV to 1\,EeV energy range. For reference, proton primary air showers reach shower maxima around 550\,g/cm$^2$ at 1\,PeV and 720\,g/cm$^2$  at 1\,EeV \cite{kampert}. Being around shower maxima is beneficial for energy resolution since shower fluctuations are smallest at shower maxima.

IceCube measures air showers on the surface with IceTop, high energy muon bundles with the in-ice detector, and both components in coincidence provided that the air shower triggers IceTop and the axis goes through the in-ice detector.

The IceTop array consists of 81 stations in its final configuration, covering an area of one square kilometer with an inter-station separation of 125\,m on average. Each station consists of two ice Cherenkov tanks separated by 10\,m. Two Digital Optical Modules (DOM) \cite{icecube} are deployed per tank. Each DOM contains a 10 inch Hamamatsu photomultiplier tube (PMT) and electronics for signal processing and readout \cite{pmt}. The two DOMs in the tank operate at different PMT gains for increased dynamic range, covering signals equivalent to more than 10$^3$ muons before saturation. An IceTop station is considered triggered when a Local Coincidence (LC) condition is satisfied initiating the readout of all waveforms and the data transfer to the IceCube Lab (ICL) at the surface. The LC condition requires that at least one of the high gain DOMs has passed the discriminator threshold and any one of the DOMs in the neighboring tank has a discriminator trigger within $\pm1\,\mu$s. DOM charges are calibrated using signals from single muons and all charges are converted to the tank and DOM independent unit of  'Vertical Equivalent Muon' (VEM) \cite{IT_detP}. Event triggers are formed in the ICL from the signals of all DOMs which have transferred data. The basic IceTop trigger for air shower physics is the IceTop Simple Majority Trigger (IceTopSMT) which requires at least 6 DOMs to have waveforms within  a sliding window of 6\,$\mu$s. IceTopSMT trigger rate is 30\,Hz.

Examples of previous analyses, using smaller IceCube configurations, can be found in reference \cite{IT26} for analysis using surface detector only, and reference \cite{IC40} for coincident events that trigger both surface and deep ice strings.

This analysis uses the surface detector only, and it is based on the data taken in the period from June 1, 2010 to May 13, 2011 when IceTop consisted of 73 stations (Fig.\ref{array}) forming a hexagon. The effective livetime of the  dataset used is 327 days. The uncertainty on livetime is less than 0.07 days which is negligible. All events which triggered at least 5 stations were processed for final analysis. This choice of selection brings the effective threshold up to 1\,PeV.

\section{\label{sim}Simulation}
Detailed simulations were used to relate measured air shower parameters to the properties of primary cosmic rays. Air showers were simulated in a wide energy range from $10^{5}$\,GeV to $10^{9.5}$\,GeV with CORSIKA v6990 \cite{corsika}. Showers above $10^{8}$\,GeV were 'thinned' \cite{corsikaman} to reduce computational time and storage volume. Hadronic interaction models used were SIBYLL 2.1 \cite{sibyll} for interactions with energies greater than 80\,GeV and FLUKA \cite{fluka} at lower energies. A smaller set was simulated using QGSJet-II-03 \cite{qgsjet} for systematic studies. CORSIKA atmosphere 12 was used as the simulated atmospheric model which is based on the July $1^{st}$, 1997 South Pole atmosphere with an atmospheric overburden of 692.9\,$\rm g/cm^{2}$ (680\,hPa). The snow cover on top of the tanks used in simulation was the same as measured in February, 2010. Air showers were simulated with equal numbers of showers per $\sin\,\theta\, \cos\,\theta$ bin where additional $\sin\theta$ term accounts for the projected detector area. Simulated zenith range was 0 to 40 degrees. Four primary types (H, He, O, Fe) were simulated with an $E^{-1}$ differential spectrum and 42000 CORSIKA showers per primary. During the analysis, showers are reweighted by different assumed spectra. Each CORSIKA shower was re-sampled 100 times to increase statistics. Shower cores were uniformly distributed over areas larger than the detector area with an energy dependent resampling radius. Resampling radii were chosen as the largest distance possible for the shower to trigger the array. The detector response was simulated using IceCube software that simulates the entire hardware and data chain \cite{IT_detP}. Interactions of charged particles with the IceTop tanks were simulated using the GEANT4 \cite{geant4} package.

\begin{figure}[!t]
  \centering
  \includegraphics[width=3.8in]{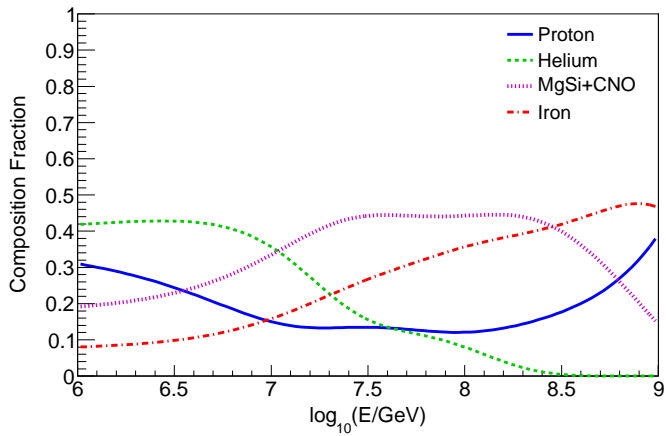}
  \caption{Fractional composition of the H4a model in 4 elemental groups. CNO and MgSi groups were combined due to lack of Mg and Si simulation.}
  \label{H4a_fr}
\end{figure}

The simulations of single primary elements were weighted by a power law spectrum, $\frac{\mathrm{d}N}{\mathrm{d}E}\propto E^{-2.7}$. For a mixed composition assumption we used the  model from reference \cite{H4a} referred to as H4a. Figure \ref{H4a_fr} shows the fractional mass composition for the H4a model. The H4a model consists of five elemental groups: H, He, CNO, MgSi and Fe. Each group has three spectral components. Each spectral component is described by a power law function with an exponential cutoff that depends on magnetic rigidity. The first component represents galactic cosmic rays from supernova remnants, the second component represents cosmic rays of unknown galactic origin, while the third component represents extra-galactic cosmic rays. Due to lack of simulation for MgSi group, oxygen simulations were weighted by the combined spectra of CNO and MgSi groups. 

\section{\label{analysis}Analysis}

\subsection{\label{basicr}Reconstructions: Direction, Core and Shower Size}
The IceTop reconstruction algorithm \cite{IT_detP} uses information from individual tanks, including location, charge and pulse time. Shower direction, core location and shower size are reconstructed by fitting the measured charges with a Lateral Distribution Function (LDF) and the signal times with a function describing the geometric shape of the shower front. The lateral distribution function is defined as:

\begin{equation}\label{eq:ldf}
S(R) = S_{ref} \left(\frac{R}{R_{ref}}\right)^{-\beta-0.303\log_{10}\left(\frac{R}{R_{ref}}\right)},
\end{equation}

\noindent where $S_{ref}$ is the shower size or signal at a reference distance $R_{ref}$ to the shower axis, and $\beta$ is the slope of the logarithmic LDF at $R_{ref}$. The shower front is described using the signal times as:

\begin{equation}\label{eq:time}
t(\mathbf{x})=t_{0}+\frac{1}{c}(\mathbf{x}-\mathbf{x}_{c})\mathbf{n}+\Delta t(R),
\end{equation}

\begin{equation}\label{eq:dtime}
\Delta t(R) = aR^{2} +b\left( 1-\exp\left(-\frac{R^{2}}{2\sigma^{2}}\right)\right),
\end{equation}

\noindent where  $a=4.823\times 10^{-4}\, \mathrm{ns/m^{2}}$, $b=19.41\, \mathrm{ns}$, $\sigma=83.5\,\mathrm{m}$, and $t(\mathbf{x})$ is the signal time of the tank at position $\mathbf{x}$, $\mathbf{x}_{c}$ is the position of shower core on the ground and $\mathbf{n}$ is the unit vector in the direction of movement of the shower.  $\Delta t(R)$ describes the deviation from the plane perpendicular to the shower axis containing $\mathbf{x}_{c}$ \cite{IT_detP}. Equations \ref{eq:ldf} and \ref{eq:time} describe the expectations for the charge and time of air shower signals. They are fitted to the measured data using a maximum likelihood method with additional terms accounting for the probability that the signal did not pass the threshold (no-hit likelihood) and that the signal was saturated (saturation likelihood, not yet implemented in \cite{IT_detP}). The shower size, S$_{125}$, is defined as the fitted value of the LDF (Eq. \ref{eq:ldf}) at a reference distance of 125\,m away from the shower axis. \\
\indent  Snow accumulates on top of IceTop tanks with time, which reduces the measured signal in a tank. To correct for this reduction, the expected signal in the likelihood fitting procedure is reduced according to:

\begin{equation}\label{snowcor}
S_{expected,\,corrected}=S_{expected}\, \exp\left(-\frac{d\,  \sec\theta}{\lambda}\right),
\end{equation}

\noindent where $d$ is the depth of snow cover on top of the tank, $\theta$ is the measured zenith angle of the shower and $\lambda =2.1\,$m is the effective attenuation length of the electromagnetic component of the shower in the snow. (See details in appendix)

\begin{table*}
\caption{\label{pass} Passing rates for quality cuts. The passing rates represent the percentage of events that passed the previous cut. Errors are statistical only. Simulation is based on the H4a model \cite{H4a}.}
\begin{ruledtabular}
\begin{tabular}{p{5.2cm} c c c c c}
                                                & \multicolumn{2}{c}{Experimental data}                       & \multicolumn{2}{c}{Simulation} \\
                                   \textbf{Cut} & \textbf{Passing rate}    & \textbf{Cumulative} & \textbf{Passing rate}        &  \textbf{Cumulative} \\

5 or more stations triggered,
 $\log_{10}(S_{125})>0.0$, $\cos\,\theta\geq0.8$     &	100		     $\%$	&			         	&	100	$\%$		&		       \\
Geometric containment	             &	58.5	$\%$	&	58.5	 $\%$	&	56.9	$\pm$	0.3 $\%$&	56.9	$\pm$	0.3 $\%$	\\
Loudest station not on edge	     &	96.6	$\%$	&	56.6	 $\%$	&	96.8	$\pm$	0.3 $\%$&	55.1	$\pm$	0.3 $\%$	\\
Largest signal $>6$\,VEM	     &	97.2	$\%$	&	55.0	 $\%$	&	98.5	$\pm$	0.4 $\%$&	54.3	$\pm$	0.3 $\%$	\\

 \end{tabular}
 \end{ruledtabular}
 \end{table*}

The core resolution of the current reconstruction method is better than 15\,m at energies around few PeV and improves to less than 8\,m at higher energies. The directional resolution is between $0.2^{\circ}-0.8^{\circ}$, depending on energy and zenith.

\begin{figure}[!t]
  \centering
  \includegraphics[width=3.6in]{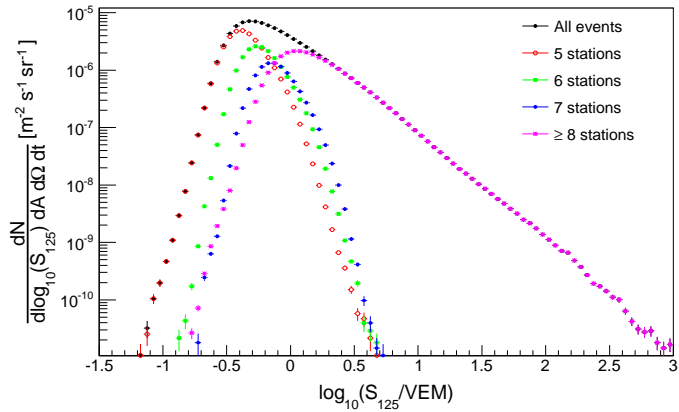}
  \caption{S$_{125}$ spectrum.  Different colors represent event selection by number of triggered stations. All cuts from \ref{cuts}, with the exception of $\log_{10}(S_{125})\geq0.0$, were applied.}
  \label{FS125}
\end{figure}

\subsection{\label{cuts}Event selection}

To improve general quality of reconstructions and to stay within the simulated zenith range, the following cuts were applied to the simulated and the experimental data:

\begin{enumerate}
\item Events must trigger at least 5 stations and with reconstruction fits converged.
\item Events must have $\log_{10}(S_{125})\geq0.0$.
\item Events must have a zenith angle with $\cos\,\theta \geq 0.8$.
\item Reconstructed cores must be within the geometric boundary shown in Fig.\ref{array}.
\item Events with the largest signal in a station on the edge of the array are rejected.
\item Events in which no station has a signal greater than 6\,VEM are rejected.
\end{enumerate}

\noindent Cut 1 was applied to select events with at least 5 stations triggered that have better reconstruction quality compared to 3 or 4 station events, while cut 2 was applied to stay above the threshold. Cut 3 was applied to stay within the simulated zenith range of $\cos\,\theta\geq0.77$. The cuts 5 and 6 were introduced to reduce the migration of high energy showers that fall outside the geometric containment but still trigger a large number of stations and get reconstructed within the containment area. The passing rates for these cuts in simulation and the experimental data are shown in Table \ref{pass}. In total, 12,253,649 events passed these quality cuts above $\log_{10}(E/\mathrm{GeV})=6.2$. Figure \ref{FS125} shows the shower size spectra for the full data sample for different numbers of triggered stations. The value of $S_{125}$ increases with the number of triggered stations, which is proportional to the primary energy. 

\begin{figure}[!t]
  \centering
  \includegraphics[width=3.4in]{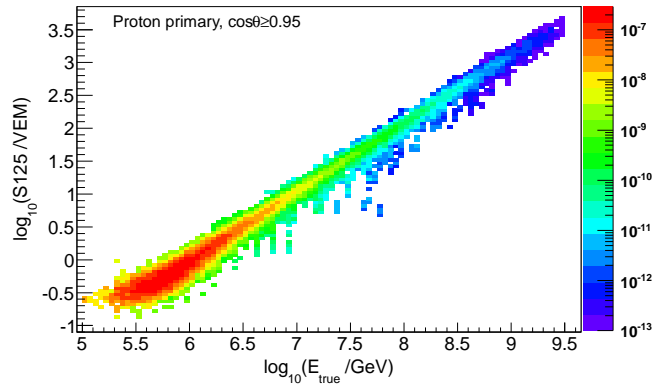}
  \caption{log$_{10}(S_{125})$ vs. log$_{10}(E_{true})$ scatter plot for proton primary simulation with cos $\theta\geq 0.95$, weighted by a flux model $\frac{dN}{dE}\propto E^{-2.7}$. }
  \label{S125Et}
\end{figure}

\begin{figure}[!t]
  \centering
  \includegraphics[width=3.6in]{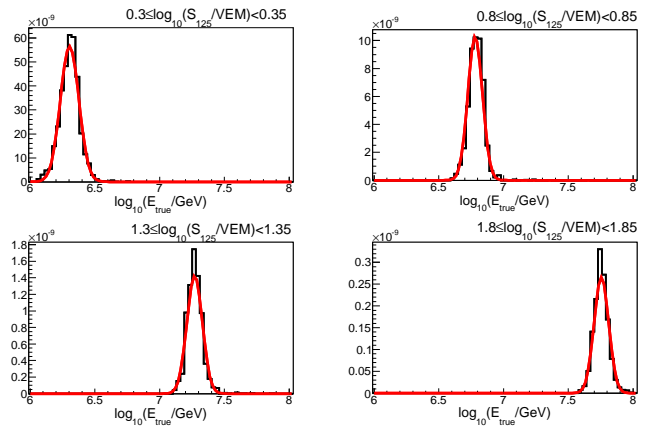}
  \caption{Examples of the true energy distributions for four $S_{125}$ slices fitted by Gaussian functions for comparison. The events were simulated using pure protons with $\frac{dN}{dE}\propto E^{-2.7}$. Shower size for the four examples are: $0.3<\log_{10}(S_{125})\leq 0.35$, $0.8<\log_{10}(S_{125})\leq 0.85$,  $1.3<\log_{10}(S_{125})\leq 1.35$,  $1.8<\log_{10}(S_{125})\leq 1.85$. The zenith range is $\cos\,\theta\geq0.95$.}
  \label{Gaus}
\end{figure}

\subsection{\label{energysec}Energy estimation method}
To estimate the energy of the primary cosmic ray, we use the relationship between the shower size $S_{125}$ and the true primary energy, $E_{true}$, from simulation. This relationship depends on the mass of the primary particle and the zenith angle of the air shower. Figure \ref{S125Et} shows a 2-dimensional histogram of the $\log_{10}(S_{125})$ vs $\log_{10}(E_{true})$ for simulated protons weighted by a flux model $\frac{dN}{dE}\propto E^{-2.7}$. For a given zenith bin we slice the distribution shown in Fig.\ref{S125Et} in 0.05 bins of log$_{10}(S_{125})$ and plot the distributions of true energy for each bin (Fig.\ref{Gaus}). We fit each energy distribution with a gaussian and use the fitted mean as the energy estimate for the given bin of $\log_{10}(S_{125})$. The relationship between log$_{10}(S_{125})$ bin and the fitted mean, log$_{10}(E_{true})$ is:

\begin{equation}\label{eq:e_conv}
\log_{10}(E) = p_{1} \log_{10}(S_{125}) + p_{0}.
\end{equation}

\begin{figure}[!t]
\begin{center}
  \subfigure[ Proton.]{
  \includegraphics[width=3.6in]{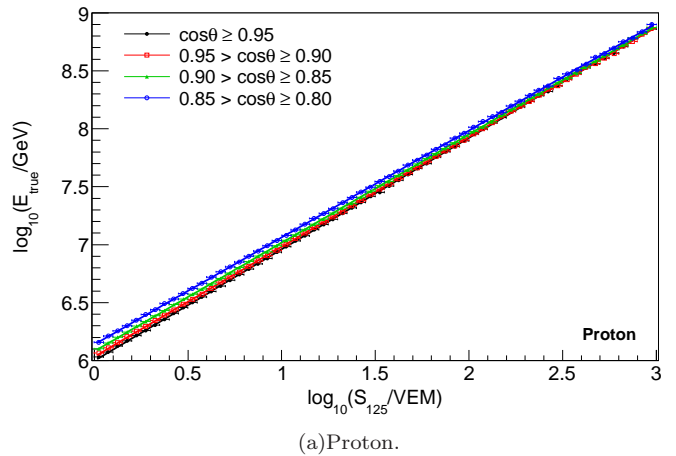}
  \label{EestP}
  } 
  \subfigure[ Iron.]{
  \includegraphics[width=3.6in]{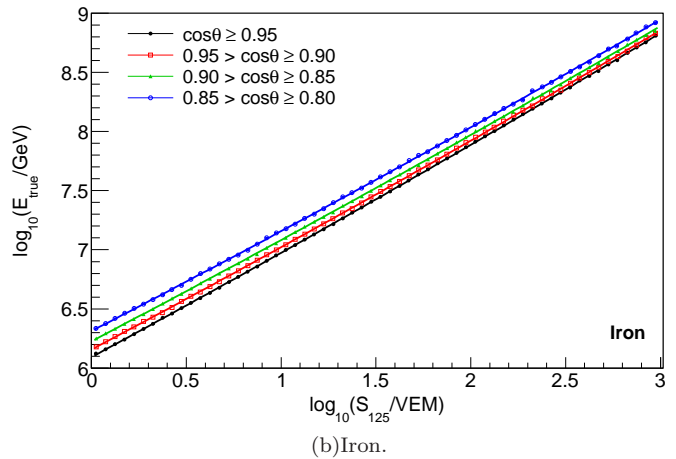}
  \label{EestFe}
  }
  \subfigure[ H4a.]{
  \includegraphics[width=3.6in]{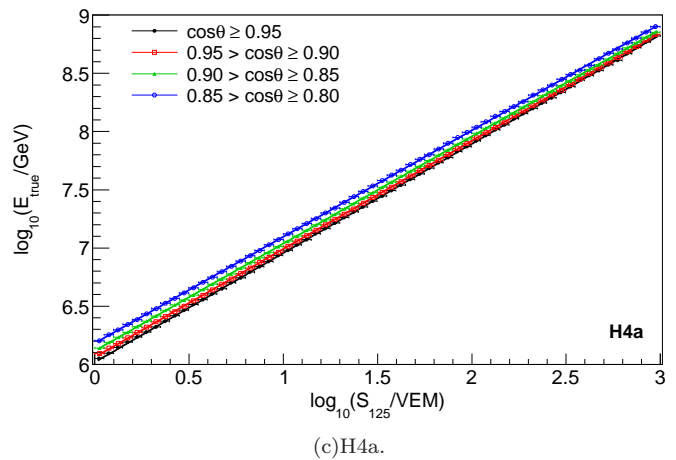}
  \label{EestH4a}
  }

\end{center}
\caption{$S_{125}$-to-$E_{true}$  relations in four zenith ranges for three composition assumptions.}
\label{EestAll}
\end{figure}

\noindent The parameters $p_{1}$ and $p_{0}$ depend on the composition assumption, the zenith angle bin and the spectral index. Table \ref{E_fits} shows the fit parameters for pure proton, pure iron and mixed H4a compositions in four zenith ranges. Energy conversion functions are calculated for each primary mass in four $\cos\,\theta$ bins: $0.80\leq\cos\,\theta<0.85, 0.85\leq\cos\,\theta<0.90, 0.90\leq\cos\,\theta<0.95$, $\cos\,\theta\geq 0.95$. In addition to four single element compositions, the mixed composition model described in the previous section was used. For each composition assumption we get a set of energy estimators as shown in Fig. \ref{EestAll} for pure proton, pure iron and the H4a model assumptions. When showing spectra for a given zenith range and assumed composition, the energy was estimated using Eq. \ref{eq:e_conv} with appropriate parameters. 

\begin{table}
\caption{\label{E_fits} Fit parameters for Eq. \ref{eq:e_conv} for three composition assumptions in four zenith ranges.}
\begin{ruledtabular}
\begin{tabular}{l c c c }                                                
\textbf{Composition} &\textbf{Zenith range}       &   \textbf{p$_{0}$} &  \textbf{p$_{1}$}  \\ \hline
\multirow{4}{*}{Proton} &   $\cos\,\theta\geq0.95$        &    5.998           &  0.962      \\ 
                        &    $0.95>\cos\,\theta\geq0.90$   &    6.034           &  0.948     \\
                        &    $0.90>\cos\,\theta\geq0.85$   &    6.081           &  0.936     \\
                        &    $0.85>\cos\,\theta\geq0.80$   &    6.139           &  0.923     \\ \hline
                                         
\multirow{4}{*}{Iron}   & $\cos\,\theta\geq0.95$        &    6.069           &  0.913        \\ 
                        & $0.95>\cos\,\theta\geq0.90$   &    6.130           &  0.900        \\
                        & $0.90>\cos\,\theta\geq0.85$   &    6.202           &  0.888        \\
                        & $0.85>\cos\,\theta\geq0.80$   &    6.288           &  0.878        \\ \hline
                                               
\multirow{4}{*}{H4a}    & $\cos\,\theta\geq0.95$  &    6.018                &    0.938      \\ 
                        & $0.95>\cos\,\theta\geq0.90$   &   6.062           &    0.929      \\
                        & $0.90>\cos\,\theta\geq0.85$   &   6.117           &    0.921      \\
                        & $0.85>\cos\,\theta\geq0.80$   &   6.182           &    0.914      \\
\end{tabular}
\end{ruledtabular}
\end{table}

\begin{figure}[!t]
  \centering
  \includegraphics[width=3.8in]{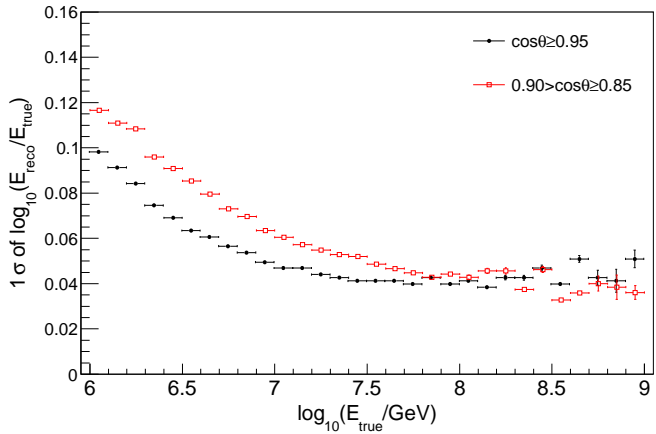}
  \caption{Energy resolution as a function of true energy for H4a assumption in two zenith bins.}
  \label{Eres}
\end{figure}

Figure \ref{Eres} shows the energy resolution defined as one sigma of the distribution of $\log_{10}(E_{reco}) - \log_{10}(E_{true})$, for a given primary and zenith bin, as a function of the true energy. Above 2\,PeV the resolution is better than 0.1 in $\log_{10}(E)$. \\

\subsection{Flux derivation}

\begin{figure}[!t]
  \centering
  \includegraphics[width=3.8in]{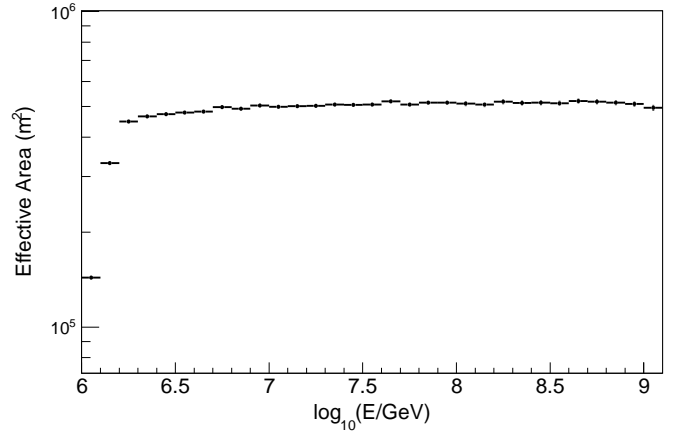}
  \caption{Effective area for H4a composition with $\cos\,\theta\geq0.8$.}
  \label{EffA}
\end{figure}

 The flux is calculated for different composition assumptions and zenith ranges according to the following definition:

\begin{equation}\label{eq:flux}
  \mathrm{J}(E) = \frac{\mathrm{d}N}{\mathrm{d}E \, A_{eff}\, \Delta\Omega\, T},
\end{equation}

\noindent where $\Delta\Omega = 2\pi\, (\cos\,\theta_{min}-\cos\,\theta_{max})$ is the solid angle range, T = livetime, and $A_{eff}$ is the effective area

\begin{equation}\label{eq:effA}
      A_{eff}(E) = A_{cut} \, \frac{\cos\,\theta_{max}+\cos\,\theta_{min}}{2} \, \epsilon(E),
\end{equation}

\noindent where $A_{cut}=577,265 \,\mathrm{m^{2}}$ is the geometric containment area in Fig.\ref{array} and  $\epsilon(E)$ is the detector efficiency

\begin{equation}\label{eq:eff}
      \epsilon(E) = \frac{N_{reco}}{N_{true}}.
\end{equation}

\noindent where  $N_{reco}$ is the number of events with reconstructed energy and zenith angle within the bin, and reconstructed core contained in the IceTop fiducial area, and $N_{true}$ is the number of events with true energy and true zenith angle within the bin, and true core contained in the IceTop fiducial area (Fig.\ref{array}). Figure \ref{EffA} shows the effective area for mixed composition and $\cos\,\theta\geq0.8$. To calculate the efficiencies for a mixed composition model, single element simulations were reweighted according to the model and the mixed efficiency was calculated. Efficiencies were evaluated and applied separately for each composition assumption and each of the four zenith bins. At maximum efficiency and $\cos\,\theta\geq0.8$, the acceptance is around 640,000$\,\mathrm{m^{2}sr}$. Examples of the derived spectra for different composition assumptions in four zenith ranges can be seen in Fig.\ref{4Zenith_Spectra}.

The final spectra were derived assuming the H4a model and averaging over the full zenith range $\cos\,\theta\geq 0.8$. The spectrum was unfolded by an iterative procedure in which the spectrum derived in the previous step was used to determine the effective area and the  $S_{125}$-to-$E_{true}$ relation for the next spectrum evaluation. In case of convergence the effective area takes correctly account of migrations due to finite resolutions. In the first step the spectrum was derived assuming the H4a model. The result was fitted by the sum of three power law functions each with an exponential cutoff. The fitted spectrum, keeping the fractional contributions of the elemental groups as in the H4a model, was used in the reweighting of the simulation for the next step efficiencies and energy conversions. The spectrum derived in this first iteration step showed no significant difference to the one derived using the original H4a model meaning that the iterative unfolding converged already after one iteration. The same algorithm was applied starting with a featureless power law spectrum with an H4a composition. In this case the spectrum converged after two iterations.

\subsection{\label{syssec}Systematics}
The four main systematic uncertainties on the flux were accounted for in this analysis. When calculating different systematics, all conditions except the systematics under investigation, are kept the same.

\subsubsection{Uncertainty in VEM calibration:}
The measured charge of each IceTop tank is calibrated using the signal from atmospheric muons \cite{IT_detP}. From simulation studies a 3\% uncertainty on the charge calibration and thus on the absolute energy scale was found \cite{arne}. This uncertainty on absolute charge calibration translates into an absolute uncertainty in the signal, $S_{125}$, and consecutively in the energy. We propagate this uncertainty to primary energy and flux.  

\subsubsection{Uncertainty in snow correction:}
\indent The systematic error due to snow correction arises from the uncertainty in the correction parameter $\lambda$ in Eq. \ref{snowcor}. In the analysis we used $\lambda=$2.1\,m and the uncertainty is $\pm$0.2\,m (see appendix).  The error in $S_{125}$ is estimated from the difference between shower size spectra derived using $\lambda=1.9$\,m and $\lambda=2.3$\,m. This error is propagated to an error in energy using the $S_{125}$-to-$E_{true}$ conversion (Eq. \ref{eq:e_conv}) for the H4a composition assumption.

\begin{figure}[!t]
  \centering
  \includegraphics[width=3.6in]{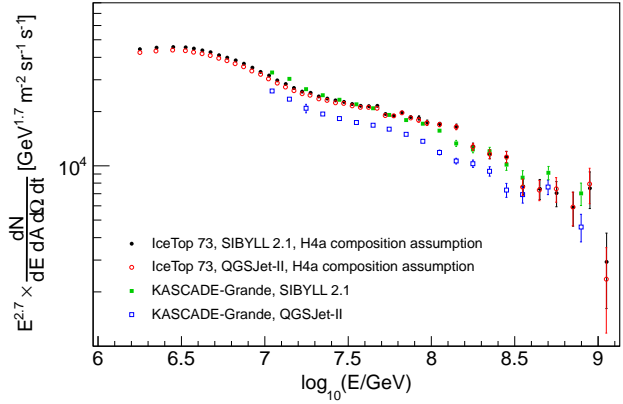}
  \caption{The difference in the spectra obtained using SIBYLL 2.1 or QGSJet-II as an interaction model for IceTop and KASCADE-Grande \cite{kascadegrande,KG_Syb}}
  \label{HadModels}
\end{figure}

\subsubsection{Difference between SIBYLL 2.1 and QGSJet-II-03:}
\indent Due to limited computational resources, only SIBYLL 2.1 and QGSJet-II-03 hadronic interaction models were used. We have chosen these two models which have also been used by other experiments, however, we are aware that they might not bracket the full uncertainty due to the interaction model. For comparison between SIBYLL 2.1 and QGSJet-II-03, the $S_{125}$-to-$E_{true}$ relations were recalculated using smaller simulated sets with QGSJet-II-03 as the interaction model. Comparison of the $S_{125}$-to-$E_{true}$ relations showed that for a given $S_{125}$, QGSJet-II-03 simulation results in lower energies compared to SIBYLL 2.1. Although we did not investigate the impact of EPOS interaction model, previous analysis \cite{IT26} showed that the difference in shower size,  between SIBYLL 2.1 and EPOS 1.99 was slightly larger compared to the difference in shower size between SIBYLL 2.1 and QGSJet-II. The largest difference in energy, between SIBYLL 2.1 and QGSJet-II, is $\Delta\log(E/\rm GeV)=0.02$ (see Table \ref{sys_t} and Fig.\ref{HadModels}). The difference in the spectra obtained using SIBYLL 2.1 or QGSJet-II-03 as an interaction model are everywhere below 4\% and thus relatively small. In Fig.\ref{HadModels} also the KASCADE-Grande results for both interaction models are shown. We note that the model differences are in that case much larger which could be due to the much lower altitude of the KASCADE-Grande detector.

\begin{figure}[!t]
\begin{center}
  \subfigure[ Proton.]{
  \includegraphics[width=3.6in]{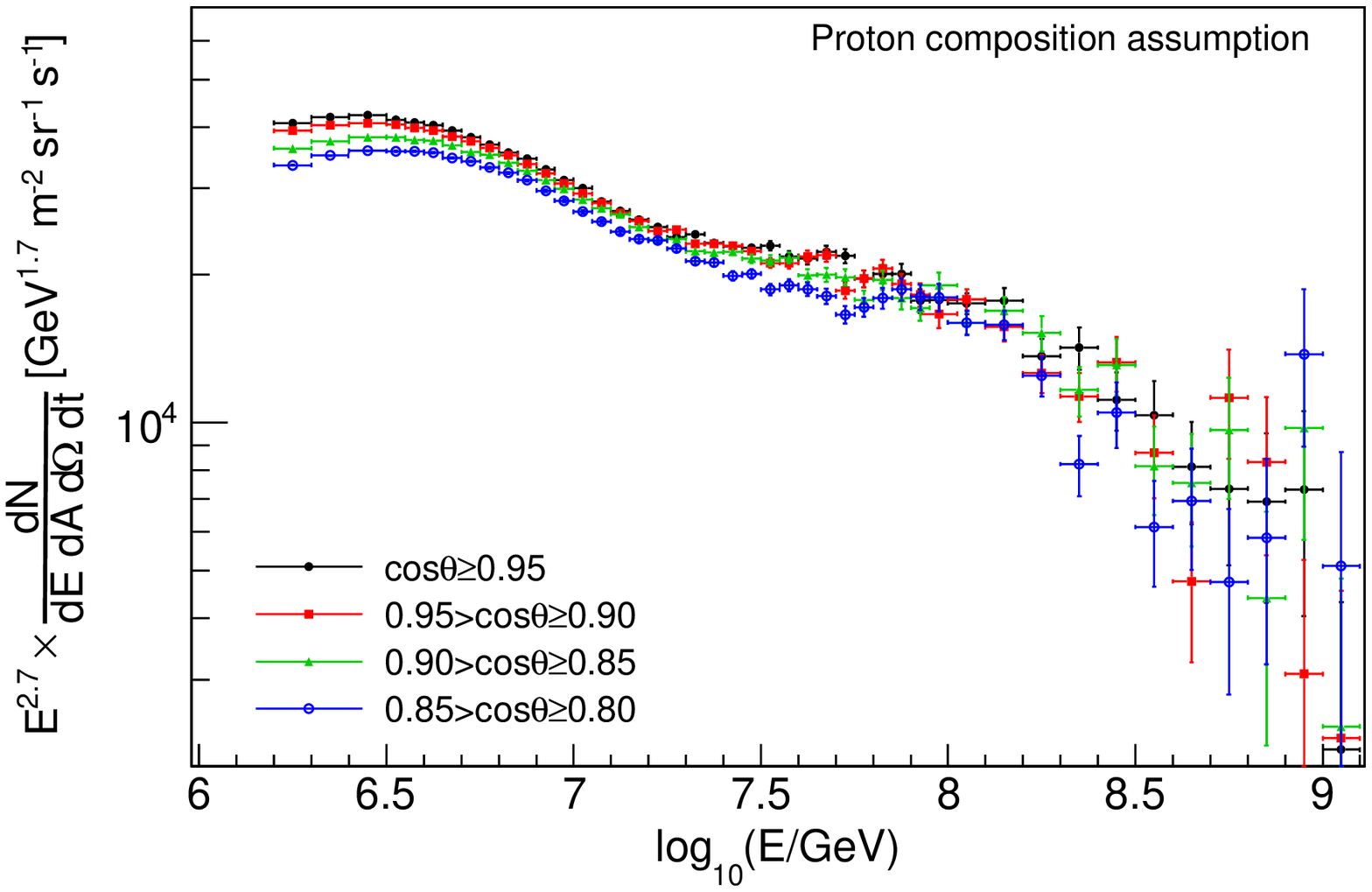}
  \label{P4Zenith}
  } 
  \subfigure[ Iron.]{
  \includegraphics[width=3.6in]{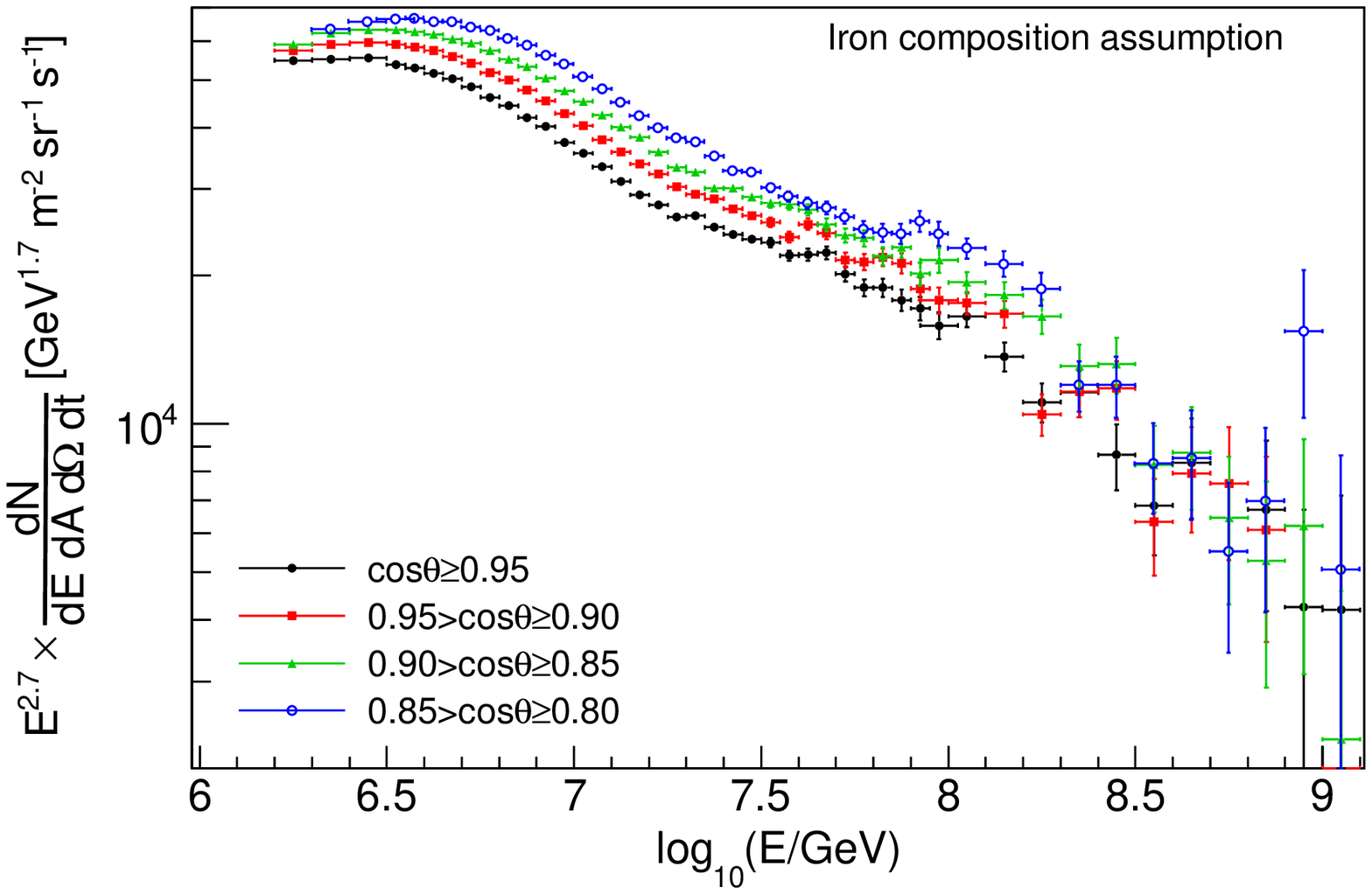}
  \label{Fe4Zenith}
  }
  \subfigure[ H4a.]{
  \includegraphics[width=3.6in]{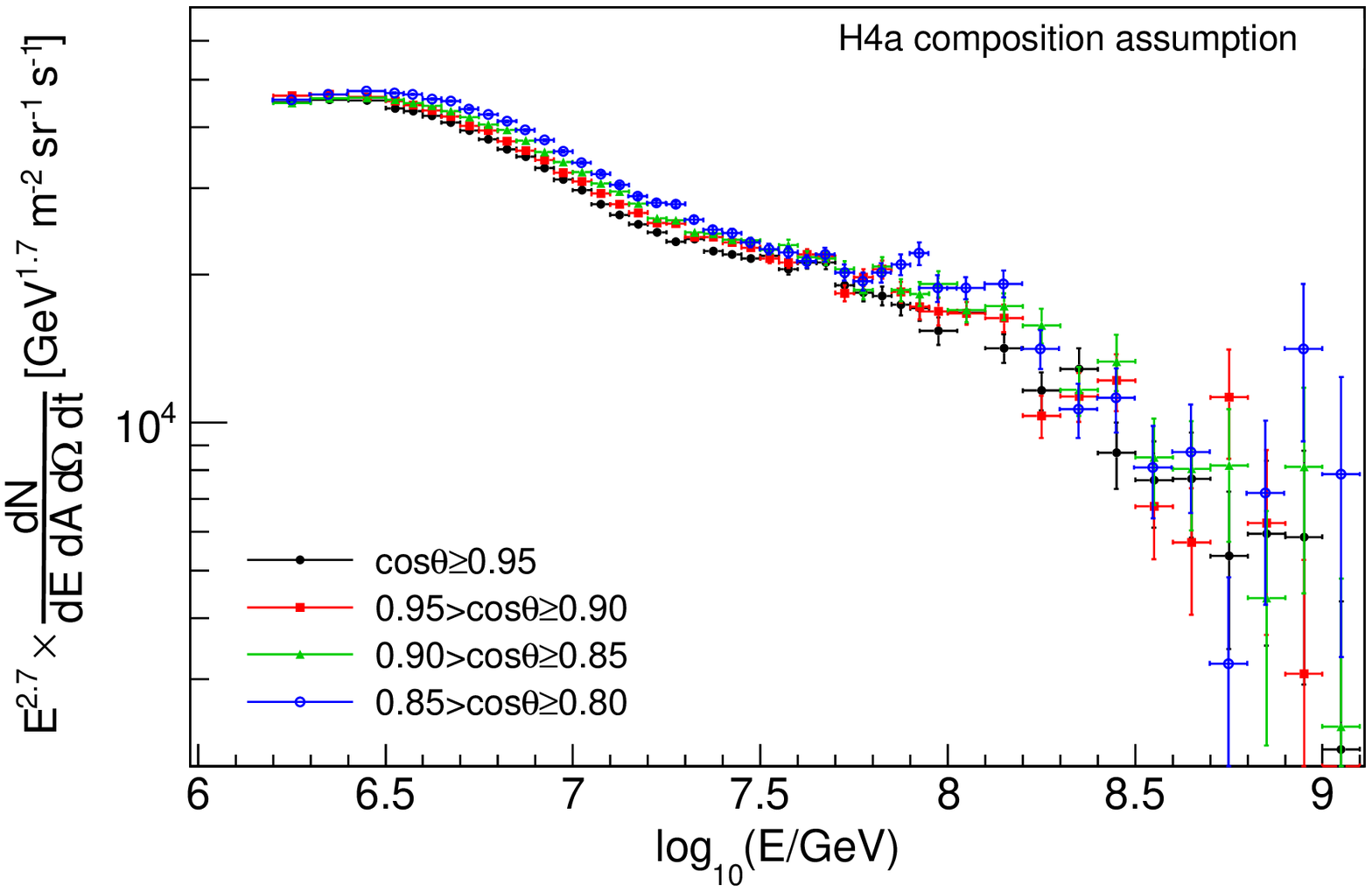}
  \label{H4a4Zenith}
  }
\end{center}
\caption{Cosmic ray energy spectrum for 3 composition assumptions and 4 zenith ranges.}
\label{4Zenith_Spectra}
\end{figure}

\subsubsection{Uncertainty and composition dependence:}
\begin{figure}[!t]
  \centering
  \includegraphics[width=3.6in]{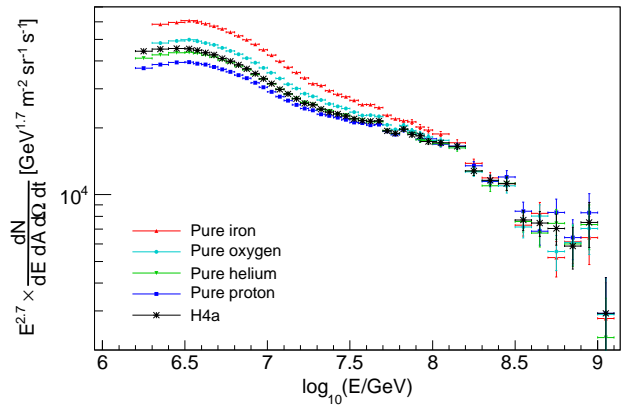}
  \caption{IceTop 73 cosmic ray energy spectrum with 5 composition assumptions and cos $\theta\geq 0.80$.}
  \label{Models}
\end{figure}

\begin{figure}[!t]
  \centering
  \includegraphics[width=3.6in]{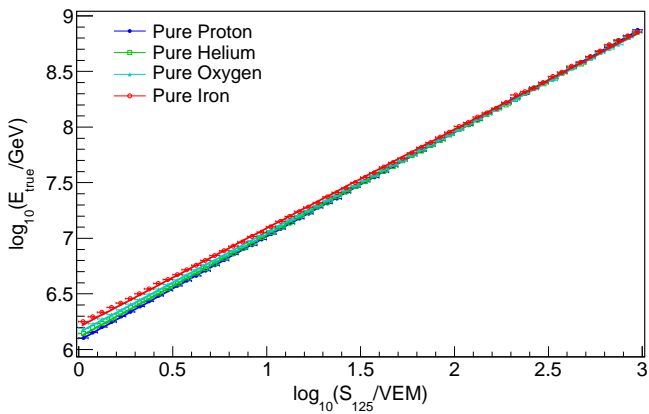}
  \caption{$S_{125}$-to-$E_{true}$  relations for four composition assumptions in $0.85\leq$ cos $\theta<0.90$ zenith range.}
  \label{Eest4}
\end{figure}

\indent The method used in this analysis requires a predefined composition assumption to translate the measured $S_{125}$ spectrum to the primary energy spectrum. Five models were tried: pure proton, pure helium, pure oxygen, pure iron and a mixed composition, H4a. Figure \ref{Models} shows the IceTop 73 spectrum with 5 composition assumptions. As shown in Fig.\ref{Eest4}, at energies above 100\,PeV the relationship  between $S_{125}$ and primary energy for different composition assumptions start to converge and cross. As a result, the spectrum measurement between 100\,PeV and 1\,EeV is relatively mass independent.

Assuming that the cosmic ray directions are isotropically distributed, the measurement of the spectrum in different zenith ranges should yield the same result for each zenith bin. For a given energy, protons or light nuclei penetrate deeper into the atmosphere compared to heavy nuclei like iron. Heavy nuclei start to interact higher in the atmosphere and showers will be at a different stage of development at the detector level compared to the light nuclei. When looking at large zenith angle events, one effectively increases the amount of atmosphere that showers need to traverse to get to the detector. This information is sensitive to composition. 

Reconstruction of the experimental data assuming pure proton and pure iron compositions in four zenith ranges are shown in Figs. \ref{P4Zenith} and \ref{Fe4Zenith}. It can be seen that for a pure proton assumption the most inclined spectrum ($0.80\leq$ cos $\theta<0.85$) is systematically lower than the vertical spectrum (cos $\theta\geq0.95$), in the energy range where statistics are not an issue. For a pure iron assumption it is the opposite, the inclined spectrum is systematically higher than the vertical. The correct composition has to agree in all zenith ranges and be in between pure proton and pure iron spectra for a given zenith range.

\begin{table}
\caption{\label{sys_t} List of systematic errors (per cent error on flux) in two energy bins.}
\begin{ruledtabular}
\begin{tabular}{p{3cm} p{2cm} p{2cm}}                                            
                                           &   \textbf{3\,PeV} & \textbf{30\,PeV}   \\ 
VEM calibration                            &  +4.0\% -4.2\%  &  +5.3\%  -5.3\%  \\
Snow                        	           &  +4.6\% -3.6\%  &  +6.3\%  -4.9\%  \\ 
Interaction models	                   &  -4.4\%	     &  -2.0\%           \\ 
Composition$^{*}$                           &  $\pm$7.0\%     &  $\pm$7.0\%    \\
Ground pressure                            &  +2.3\% -2.0\%  &  +0.4\%  -1.0\%  \\ 

 \end{tabular}
 \end{ruledtabular}
\begin{flushleft}\textbf{$^{*}$} Composition uncertainty is not constant with energy but the largest value was chosen as a fixed, conservative estimate.\end{flushleft}
 \end{table}

\begin{figure*}[!t]
  \centering
  \includegraphics[width=5.8in]{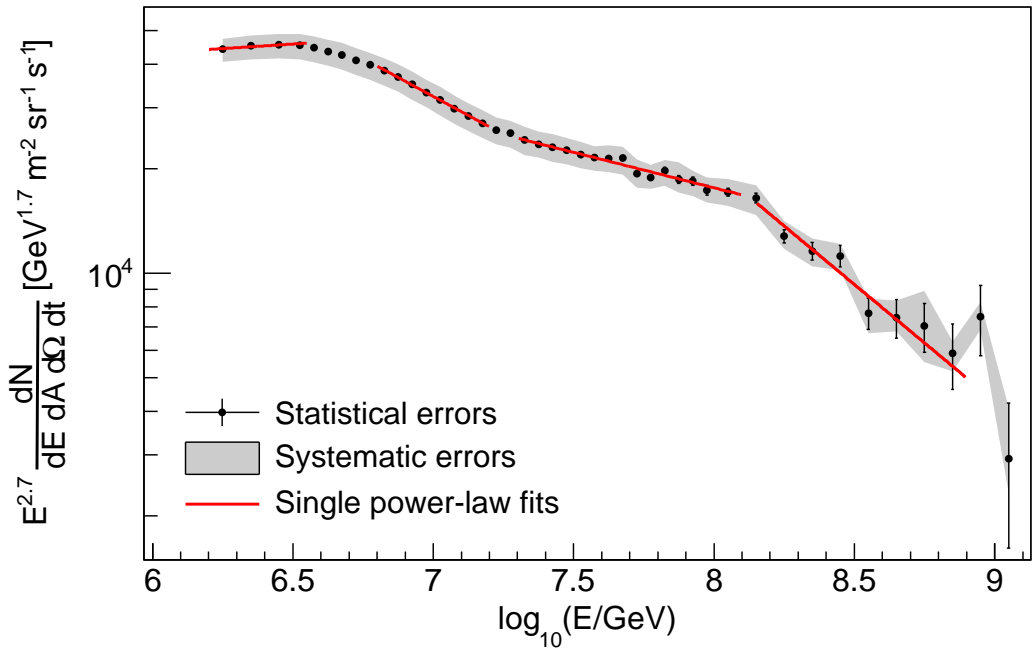}
  \caption{Spectral fits in different energy ranges.}
  \label{gamma_fit}
\end{figure*}

Four zenith spectra for a mixed, H4a, composition assumption can be seen in Fig.\ref{H4a4Zenith}. Compared to pure proton and pure iron, the mixed assumption leads to a smaller difference between vertical and inclined spectra, but still not zero. The final spectrum is determined using the H4a model in the zenith angle range cos $\theta_{min}=1.0$, and cos $\theta_{max}=0.8$. To estimate the systematic uncertainty in the all-particle energy spectrum due to composition, we use the differences for the H4a assumption between the final and the vertical (cos $\theta\geq0.95$) spectra, and the final and the most inclined ($0.80\leq$ cos $\theta<0.85$) spectra in the energy range $6.2<\log_{10}(E/\rm GeV)<7.5$ where statistical fluctuations are negligible. Although at high energies the $S_{125}$-to-$E_{true}$ relation is relatively mass independent (Figs.\ref{Models},\ref{Eest4}), the largest difference between spectra is taken as a fixed value for the error due to composition across all energies as a conservative estimate.

\subsubsection{Impact of ground pressure:}
The impact of ground pressure on the measured flux was also investigated by looking at spectra from different data samples with high (690\,hPa) and low (670\,hPa) average pressures. Changes in the flux between high and low pressure subsamples were less than $\sim$2\% and the variations averaged out when taking the full year of data with an average pressure of 680\,hPa. The simulated pressure was also 680\,hPa.  

The comparison of these four systematic errors can be seen in Table \ref{sys_t}.

\section{Results and Discussion}

\begin{figure*}[!t]
  \centering
  \includegraphics[width=5.8in]{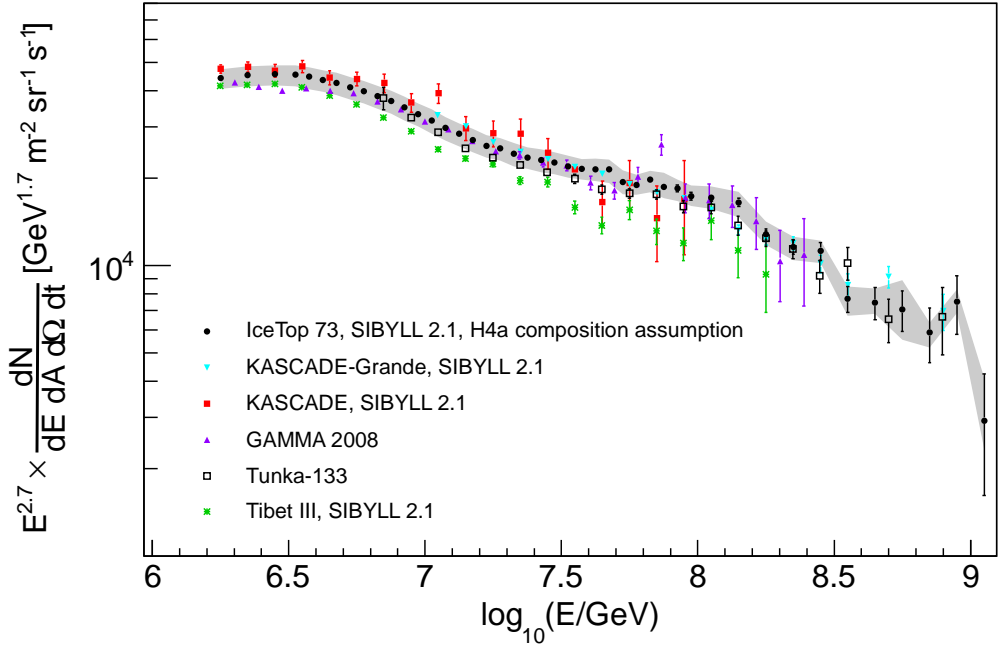}
  \caption{IceTop 73 spectrum in comparison to other recent (within the last 6 years) experiments: KASCADE \cite{kascade}, KASCADE-Grande \cite{kascadegrande}, Tunka-133 \cite{tunka}, GAMMA \cite{gamma}, Tibet \cite{tibet}. Errors are statistical only.}
  \label{Comp}
\end{figure*}

The final spectrum is shown in Fig.\ref{gamma_fit}. The IceTop shower size parameter, $S_{125}$, is calibrated against the true primary energy using the H4a composition model as an input to our simulations. We observe that, beyond our systematics, the all-particle cosmic-ray energy spectrum does not follow a single power law above the knee ($4.4\pm0.4$\,PeV), but shows significant structure.  The final spectrum was fitted by simple power law functions of the form

\begin{equation}\label{eq:fit}
  \frac{\mathrm{d}N}{\mathrm{d}\ln E\, \mathrm{d}A\, \mathrm{d}\Omega\, \mathrm{d}t} = I_{0}  \left(\frac{E}{1\,\mathrm{GeV}}\right)^{-\gamma+1},
\end{equation}

\noindent in four different energy ranges. The fits to the spectrum are shown in Fig.\ref{gamma_fit} and their parameters in Table \ref{G_fits}.  The $\chi^2$ values have been derived using the statistical errors only which may underestimate the actual uncertainties. The first interval is not well fitted which could be caused by bin-to-bin systematic uncertainties or by a wrong assumption about the fitting function. The obtained slope parameter, however,  is in good agreement with those obtained by other experiments. To estimate the systematic errors on fitted parameters, the same fitting procedure was applied to the different spectra from the previous section where the spectra changed by varying each of the systematics. The differences in fitted parameters due to four systematics (VEM calibration, snow correction, composition, interaction model) were used as the systematic errors and were added in quadrature.  

\begin{table}
\caption{\label{G_fits} Results of the fits with a power law function (Eq. \ref{eq:fit}) to the final spectrum with the H4a model for composition assumption. Energy range is in $\log_{10}(E/\mathrm{GeV})$ and I$_0$ is in $\mathrm{m^{-2}sr^{-1}s^{-1}}$.}
\begin{ruledtabular}
\begin{tabular}{p{1.4cm} p{2.6cm} p{2.9cm}  r }                       
\textbf{E range}                     &   \textbf{I$_0 \pm\mathrm{stat.}$} & \textbf{$\gamma \pm \mathrm{stat.} \pm \mathrm{sys.}$}  &\textbf{$\chi^{2}/ndf$}\\ 
6.20--6.55                           &  $(2.107\pm0.06)\times 10^{4}$     & 2.648  $\pm$ 0.002  $\pm$  0.06  & 206/2 \\ 
6.80--7.20                           &  $(3.739\pm0.34)\times 10^{7}$     & 3.138  $\pm$ 0.006  $\pm$  0.03  & 14/6   \\
7.30--8.00                           &  $(7.494\pm1.29)\times 10^{5}$     & 2.903  $\pm$ 0.010  $\pm$  0.03  & 19/12  \\
8.15--8.90 	                     &  $(4.952\pm1.65)\times 10^{9}$     & 3.374  $\pm$ 0.069  $\pm$  0.08  & 8/6    \\ 
\end{tabular}
\end{ruledtabular}
\end{table}

The differential spectral index before the knee is $-2.63\pm0.01\pm0.06$, and changes smoothly between 4 to 7\,PeV ($\log_{10}(E/\mathrm{GeV})=6.6-6.85$) to $-3.13\pm0.01\pm0.03$. Another break is observed at around $18\pm2$\,PeV ($\log_{10}(E/\mathrm{GeV})=7.3$), above which the spectrum hardens with a differential spectral index of $-2.91\pm0.01\pm0.03$. The break points in the spectrum are defined as the intersection of the fitted power law functions. A sharp fall is observed beyond $130\pm30$\,PeV ($\log_{10}(E/\mathrm{GeV})=8.1$) with a differential spectral index of $-3.37\pm0.08\pm0.08$. Above 100\,PeV, the measurement of the spectrum is relatively mass independent as can be seen in Fig.\ref{Models}. 

The significance that the observed spectra cannot be described by one or two power law functions only, can be seen in the differences of the fitted slopes and their uncertainties in Table \ref{G_fits}. The difference in the slopes between the first and the second, the second and third and the third and fourth energy ranges are 7\,$\sigma$, 5.5\,$\sigma$ and 4\,$\sigma$, respectively. In addition, we studied the extrapolations of the fits in one energy range to the energy ranges above the fitted one. For example, if we extrapolate the fit in the second energy range  (with $\gamma=3.14$)  we expect to see above that energy range about  124800 events while we observe 139880. The difference is about $43 \sqrt{N}$ showing the incompatibility of the data with the assumption that the spectrum above the knee can be fitted by only one power law function. Similarly, the extrapolation of the fit in the third energy range to energies above yields 4213 expected events while 3673 are observed. The discrepancy is about $8 \sqrt{N}$.

\begin{figure}[!t]
  \centering
  \includegraphics[width=3.8in]{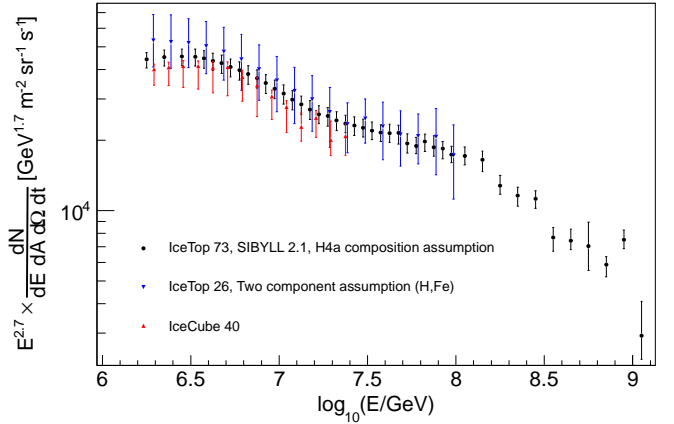}
  \caption{IceTop 73 spectrum in comparison to previous IceCube results, \cite{IT26,IC40}. Errors bars are systematic errors added in quadrature.}
  \label{comp_ic_only}
\end{figure}

We compare the IceTop-73 result with previous IceCube results in Fig.\ref{comp_ic_only}, and other, relatively recent, experiments in the PeV to EeV energy range in Fig.\ref{Comp}. Comparison of IceCube results to older experiments can be found in \cite{IT26} and \cite{IC40}.

The IceTop-73 result agrees within systematics both with IceTop-26 \cite{IT26} and IceCube-40 \cite{IC40} results. The major differences between the IceTop-73 and previous analyses are the assumed composition model, different snow treatment, improvements in the reconstruction and simulation codes, a larger detector and a longer data taking period.  

This result agrees relatively well with Tunka \cite{tunka} and GAMMA \cite{gamma} results, except for the spike around 60\,PeV in the Gamma spectrum which we can not confirm. The agreement with KASCADE \cite{kascade} and KASCADE-Grande \cite{kascadegrande} results is within systematic errors. All experiments show similar structure in the spectra, however, the breaks at 18\,PeV ($\log_{10}(E/\mathrm{GeV})=7.3$) and 130\,PeV ($\log_{10}(E/\mathrm{GeV})=8.1$) appear to be most significant in the IceTop 73 result. 

\section{Summary and Outlook}
In summary, we have obtained a measurement of the cosmic-ray spectrum with a resolution of 25\% around 2\,PeV which improves to 12\% above 10\,PeV, using one year of data from the nearly complete IceTop array. The result obtained assumes a mixed composition based on the H4a model \cite{H4a}. The hardening of the spectrum around 20\,PeV and steepening around 130\,PeV is a clear signature of the spectrum and can not be attributed to any of the systematics or detector artefacts. Thus, any model trying to explain the acceleration and propagation of cosmic rays needs to reproduce these features. 

The potential for obtaining further and more detailed information about the primary cosmic-ray spectrum with IceCube is not yet fully exhausted. Analysis of coincident events over the same period as this analysis is currently underway, including improved treatment of photon propagation in the ice and correcting for seasonal variations to be able to use the full year of data without extra systematics.  The acceptance can be more than doubled by using the full IceCube as a cosmic-ray detector and extending the zenith angle range to greater than 60 degrees.  This can be done for showers with cores in IceTop, for showers with cores through the deep detector and for an energy-dependent fraction of coincident events. Use of several independent and complementary measures of spectrum and composition to cross-calibrate the different approaches will place an important consistency constraint on the conclusions. Finally, the use of single station hits and 3-station events, including several more closely spaced tanks deployed in the final construction season of IceCube, will decrease the threshold for the analysis by an order of magnitude, to give some overlap with direct measurements.

\begin{acknowledgments}
We acknowledge the support from the following agencies:
U.S. National Science Foundation-Office of Polar Programs,
U.S. National Science Foundation-Physics Division,
University of Wisconsin Alumni Research Foundation,
the Grid Laboratory Of Wisconsin (GLOW) grid infrastructure at the University of Wisconsin - Madison, the Open Science Grid (OSG) grid infrastructure;
U.S. Department of Energy, and National Energy Research Scientific Computing Center,
the Louisiana Optical Network Initiative (LONI) grid computing resources;
Natural Sciences and Engineering Research Council of Canada,
WestGrid and Compute/Calcul Canada;
Swedish Research Council,
Swedish Polar Research Secretariat,
Swedish National Infrastructure for Computing (SNIC),
and Knut and Alice Wallenberg Foundation, Sweden;
German Ministry for Education and Research (BMBF),
Deutsche Forschungsgemeinschaft (DFG),
Helmholtz Alliance for Astroparticle Physics (HAP),
Research Department of Plasmas with Complex Interactions (Bochum), Germany;
Fund for Scientific Research (FNRS-FWO),
FWO Odysseus programme,
Flanders Institute to encourage scientific and technological research in industry (IWT),
Belgian Federal Science Policy Office (Belspo);
University of Oxford, United Kingdom;
Marsden Fund, New Zealand;
Australian Research Council;
Japan Society for Promotion of Science (JSPS);
the Swiss National Science Foundation (SNSF), Switzerland;
National Research Foundation of Korea (NRF)

\end{acknowledgments}

\appendix*

\section{\label{snow_prob}Snow}

\begin{figure}[!t]
  \centering
  \includegraphics[width=3.4in]{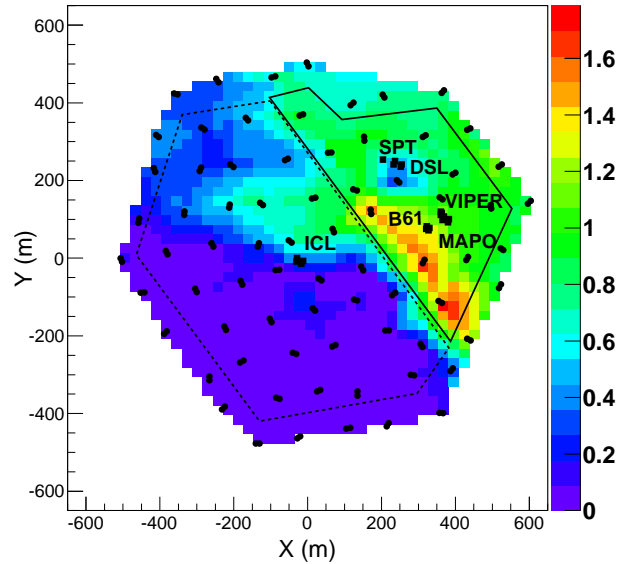}
  \caption{IceTop 73 snow cover in meters. Dashed polygon represent the 'New', less snowy part of the detector; solid polygon shows the 'Old', snowy part of the detector.}
  \label{SnowCov}
\end{figure}

\begin{figure*}[!t]
  \begin{center}
  \subfigure[No snow correction.]{
  \includegraphics[width=3.3in]{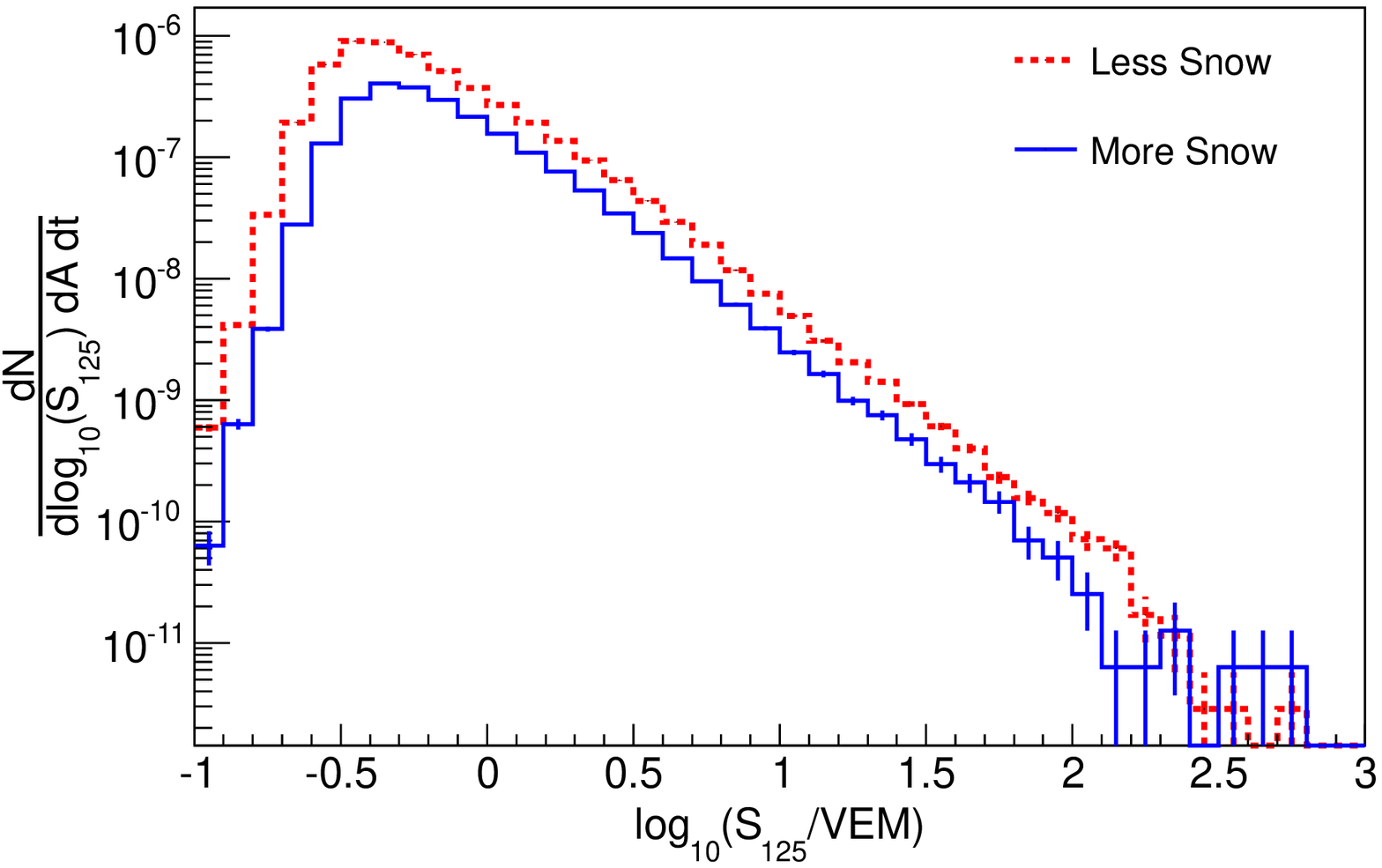}
  \label{OldNew}
  }
  \subfigure[With snow correction.]{
  \includegraphics[width=3.4in]{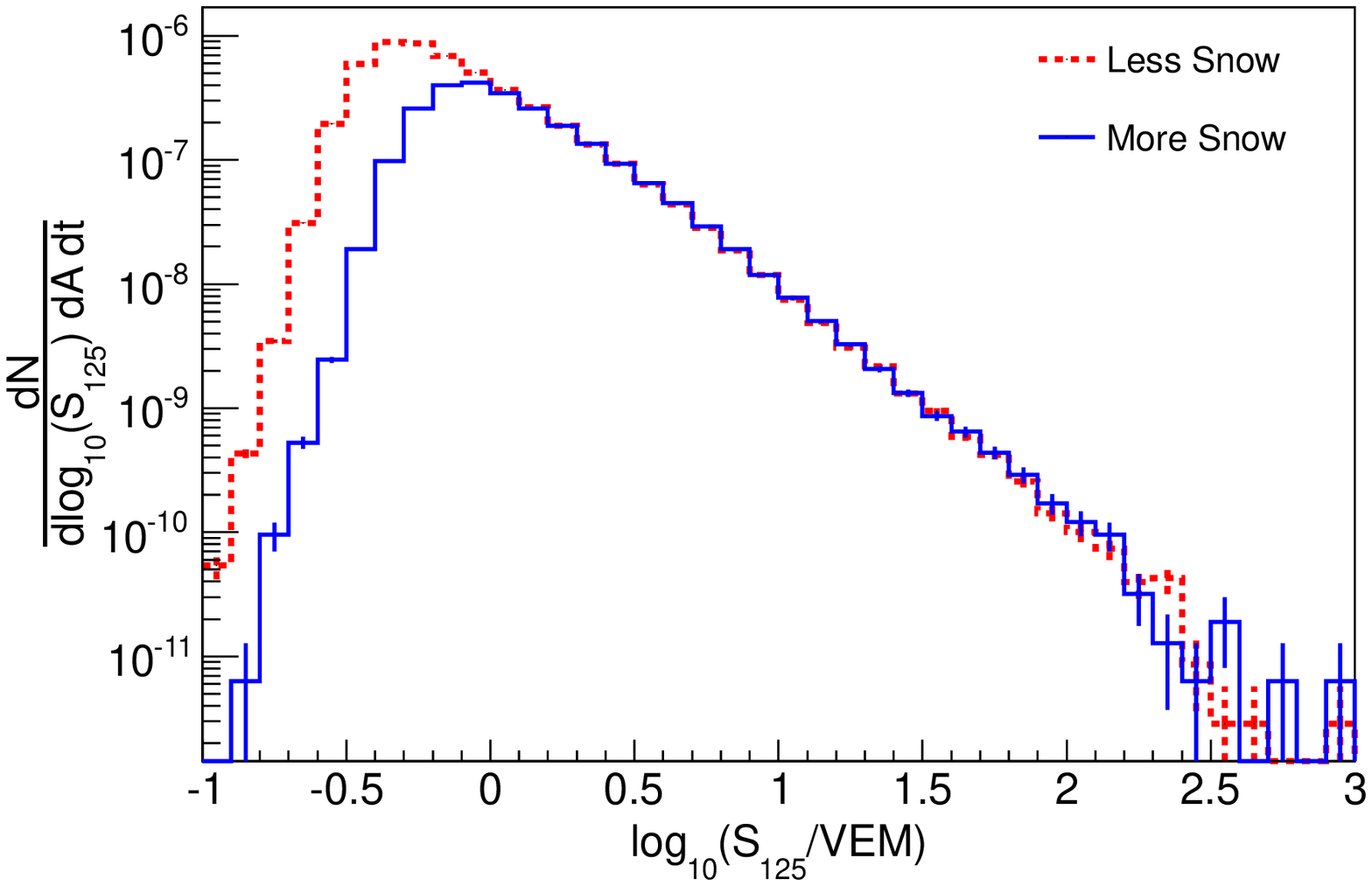}
  \label{OldNewCorr}
  }
  \end{center}
  \caption{Shower size spectra for two containment cuts weighted by their respective areas with and without snow correction.}
  \label{OldNewS125}
\end{figure*}

\begin{figure*}[ht!]
  \begin{center}
  \subfigure[No snow correction.]{
  \label{ShowerCoreNoSnow}
  \includegraphics[width=3.0in]{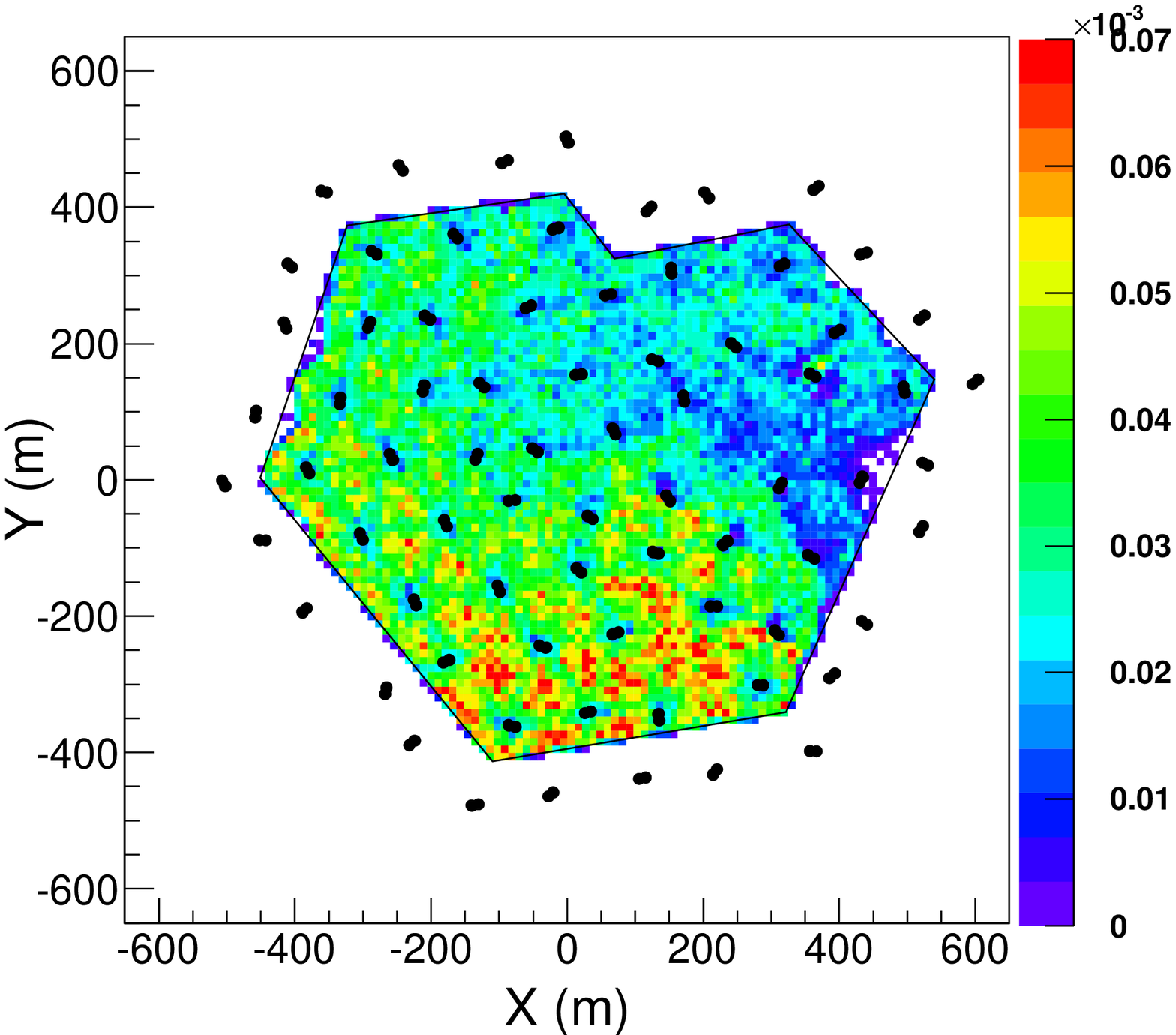}
  }
  \subfigure[With snow correction.]{
  \label{ShowerCoreSnow}
  \includegraphics[width=3.0in]{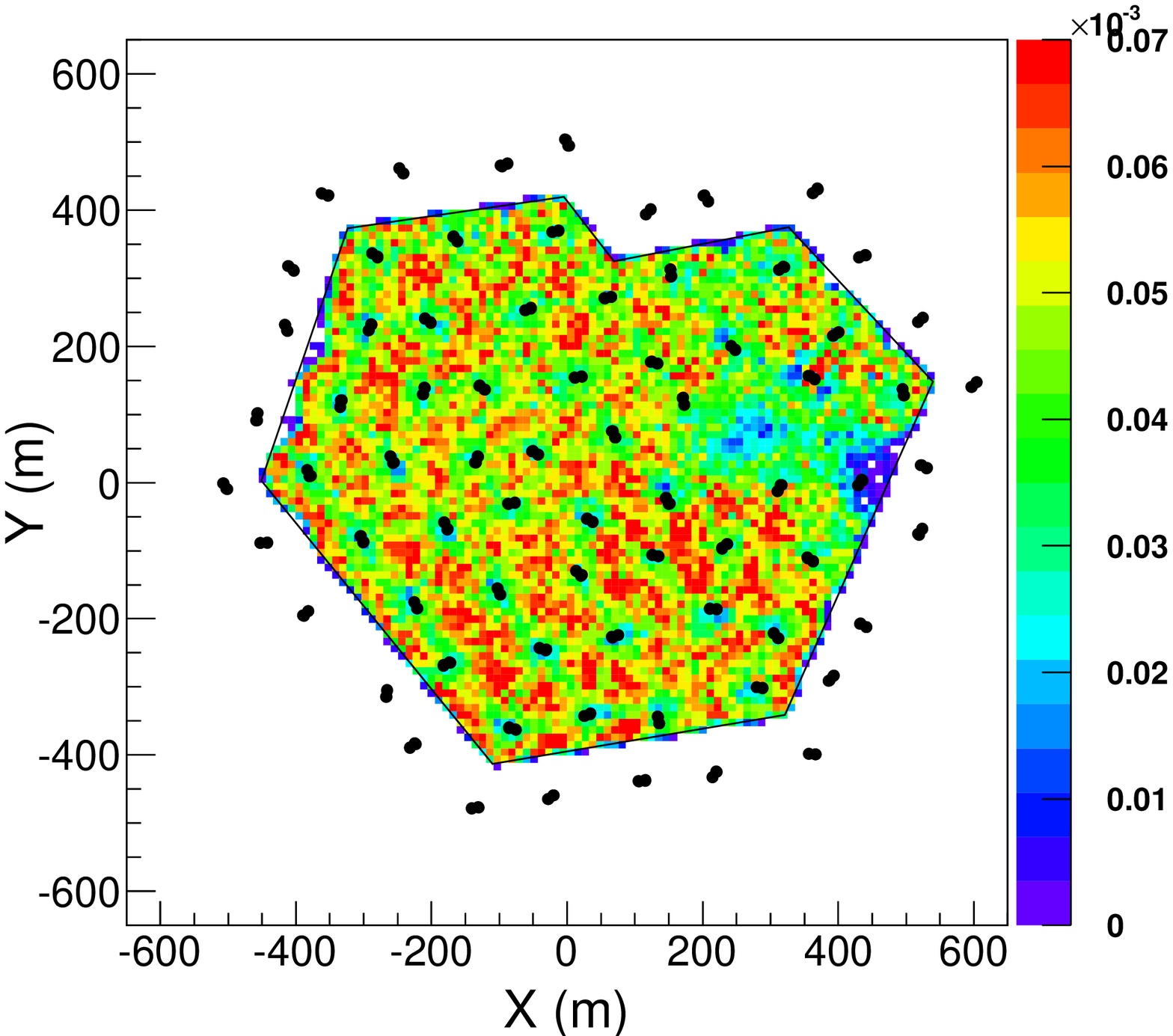}
  }
  \end{center}
  \caption{Shower core distribution for one month of data, $\cos\, \theta\geq0.95$, with 5 or more stations triggered, $\log_{10}(S_{125}/\mathrm{VEM})>0.0$.}
\label{ShowCores}  
\end{figure*}

Although South Pole is the driest place on Earth with little precipitation, snow can accumulate on top of tanks due to drifting. The surface topology and presence of nearby buildings can affect this process. IceTop records signals from tanks that come in coincidence with a signal from the neighboring tank at the same station. This signal is dominated by the electromagnetic component of the air shower. Unlike muons that are highly penetrating, electrons and photons are affected by the snow. Electrons and photons can either be absorbed by snow or produce cascades. From simulation studies it was found that absorption is the dominant effect for electrons and gammas with energies less than 1\,GeV. This is likely because the ice in the tank is two radiation lengths thick and is the main target for converting photons, which dominate the electromagnetic signal. Since the main signal in IceTop tanks is due to photons and low energy electrons, snow on top of the IceTop tanks tends to reduce the signal.

\indent The effect of snow can be seen if we geometrically separate showers according to their shower core locations. Figure \ref{SnowCov} shows the IceTop 73 detector geometry with snow coverage indicated by the color scale. Two polygons represent two containment subsets. The first subset, called 'Old', represents showers that fell in the snowy part of the detector. The second subset, called 'New', represents showers that fell in the less snowy part of the detector.
Figure \ref{OldNew} shows the uncorrected shower size spectra for these two subsets weighted by their respective containment areas. Since all showers were taken during the same period of time, all atmospheric conditions, like pressure, temperature, etc. were the same for both subsets. It is clearly seen that showers that fell into the snowy part of the detector get a smaller reconstructed shower size, $S_{125}$.\\
\indent  To estimate $\lambda$, a range of possible values from 1.5\,m to 4.0\,m was used in the correction Eq. \ref{snowcor}. $\lambda$ of $2.1\pm 0.2$\,m was chosen as the value that reconciles $S_{125}$ spectra from different parts of the detector that have different snow cover. During the reconstruction process, the likelihood algorithm tries to minimize the difference between the measured signal of each tank and the signal expected from simulations. The snow correction of Eq. \ref{snowcor} is applied to reduce the expected value of signals in tanks under snow in the likelihood fitting procedure.\\
\indent  Figure \ref{OldNewCorr} shows the shower size spectra for 'New' and 'Old' containment cuts with the snow correction applied. After correction both parts of the detector give the same shower size spectra. Of course low energy showers that fell into the 'Old' detector and did not trigger but could have triggered if they had fallen into the 'New' detector will not be recovered by this correction. \\
\indent Another way to see the effect of snow is by looking at the shower core distributions (Figures \ref{ShowerCoreNoSnow} and \ref{ShowerCoreSnow}). Snow effectively lowers the measured shower size; $S_{125}$, for a given primary energy. As a result, above a certain shower size, parts of the detector with more snow cover will trigger less often because a given $S_{125}$ corresponds to a higher primary energy compared to the less snowy part. Since the flux decreases with primary energy, the snowy part of the detector will have lower rates. This can be seen as fewer reconstructed shower cores in that part of the detector (see figure \ref{ShowerCoreNoSnow}). Snow correction ensures that independent of where the shower falls in the detector, the measured shower size will correspond to the same primary energy (assuming the same mass and atmospheric conditions). \\
\indent The snow on top of the tanks is measured twice a year. In between these measurements, snow accumulation is estimated by the method described in \cite{IT_detP} which is based on the ratio of the electromagnetic to muon component of the calibration curve. This method is accurate up to 20\,cm. The snow density is  $0.35-0.4\,\rm{g/cm^{3}}$ and $\lambda=2.1$\,m corresponds to $84\,\rm{g/cm^{2}}$. The attenuation parameter $\lambda$ in Eq. \ref{snowcor} has an energy dependent behavior. In this analysis we used the average value for $\lambda=2.1$\,m but it may vary by $\pm0.2$\,m. The value of 0.2 m comes from the comparison of $S_{125}$ spectra with different energy and zenith ranges.

 \begin{table*}[H] 
 \caption{\label{data} Spectrum data}
 \begin{ruledtabular}
 \begin{tabular}{ p{3cm} p{4cm} p{8cm} }

 $\log_{10}(E/\rm{GeV})$ bin &  Number of events per bin & $\frac{\mathrm{d}N}{\mathrm{d}\ln(E)\mathrm{d}A\mathrm{d}t\mathrm{d}\Omega}\pm \mathrm{stat + syst - syst} \,(\rm{m^{-2} s^{-1} sr^{-1}})$ \\

\hline

6.20	--	6.30	&	$396.6 \times 10^4$	&	$	(	10.495	\pm	0.006	+	0.729	-	0.855	)	\times 10^{-7}	$	\\
6.30	--	6.40	&	$278.1 \times 10^4$	&	$	(	7.250	\pm	0.005	+	0.523	-	0.612	)	\times 10^{-7}	$	\\
6.40	--	6.50	&	$191.3 \times 10^4$	&	$	(	4.938	\pm	0.004	+	0.368	-	0.425	)	\times 10^{-7}	$	\\
6.50	--	6.55	&	708089    	&	$	(	3.670	\pm	0.004	+	0.286	-	0.325	)	\times 10^{-7}	$	\\
6.55	--	6.60	&	579534	        &	$	(	2.969	\pm	0.004	+	0.230	-	0.276	)	\times 10^{-7}	$	\\
6.60	--	6.65	&	469844  	&	$	(	2.382	\pm	0.003	+	0.189	-	0.216	)	\times 10^{-7}	$	\\
6.65	--	6.70	&	379797  	&	$	(	1.914	\pm	0.003	+	0.156	-	0.180	)	\times 10^{-7}	$	\\
6.70	--	6.75	&	302695  	&	$	(	1.517	\pm	0.003	+	0.125	-	0.140	)	\times 10^{-7}	$	\\
6.75	--	6.80	&	242627  	&	$	(	1.210	\pm	0.002	+	0.100	-	0.113	)	\times 10^{-7}	$	\\
6.80	--	6.85	&	192910  	&	$	(	9.582	\pm	0.022	+	0.803	-	0.929	)	\times 10^{-8}	$	\\
6.85	--	6.90	&	152793  	&	$	(	7.562	\pm	0.019	+	0.644	-	0.707	)	\times 10^{-8}	$	\\
6.90	--	6.95	&	119945  	&	$	(	5.916	\pm	0.017	+	0.517	-	0.584	)	\times 10^{-8}	$	\\
6.95	--	7.00	&	93839   	&	$	(	4.608	\pm	0.015	+	0.409	-	0.430	)	\times 10^{-8}	$	\\
7.00	--	7.05	&	73785   	&	$	(	3.609	\pm	0.013	+	0.323	-	0.358	)	\times 10^{-8}	$	\\
7.05	--	7.10	&	57413   	&	$	(	2.798	\pm	0.012	+	0.252	-	0.267	)	\times 10^{-8}	$	\\
7.10	--	7.15	&	45112   	&	$	(	2.193	\pm	0.010	+	0.189	-	0.211	)	\times 10^{-8}	$	\\
7.15	--	7.20	&	35386	        &	$	(	1.717	\pm	0.009	+	0.156	-	0.161	)	\times 10^{-8}	$	\\
7.20	--	7.25	&	27813   	&	$	(	1.347	\pm	0.008	+	0.119	-	0.119	)	\times 10^{-8}	$	\\
7.25	--	7.30	&	22515    	&	$	(	1.088	\pm	0.007	+	0.092	-	0.103	)	\times 10^{-8}	$	\\
7.30	--	7.35	&	17722    	&	$	(	8.554	\pm	0.064	+	0.777	-	0.814	)	\times 10^{-9}	$	\\
7.35	--	7.40	&	14175	        &	$	(	6.835	\pm	0.057	+	0.578	-	0.588	)	\times 10^{-9}	$	\\
7.40	--	7.45	&	11416   	&	$	(	5.502	\pm	0.051	+	0.499	-	0.511	)	\times 10^{-9}	$	\\
7.45	--	7.50	&	9198    	&	$	(	4.433	\pm	0.046	+	0.383	-	0.393	)	\times 10^{-9}	$	\\
7.50	--	7.55	&	7351    	&	$	(	3.543	\pm	0.041	+	0.310	-	0.306	)	\times 10^{-9}	$	\\
7.55	--	7.60	&	5925     	&	$	(	2.856	\pm	0.037	+	0.225	-	0.237	)	\times 10^{-9}	$	\\
7.60	--	7.65	&	4844    	&	$	(	2.335	\pm	0.033	+	0.214	-	0.205	)	\times 10^{-9}	$	\\
7.65	--	7.70	&	3994    	&	$	(	1.925	\pm	0.030	+	0.150	-	0.200	)	\times 10^{-9}	$	\\
7.70	--	7.75	&	2965    	&	$	(	1.429	\pm	0.026	+	0.137	-	0.130	)	\times 10^{-9}	$	\\
7.75	--	7.80	&	2377    	&	$	(	1.146	\pm	0.023	+	0.100	-	0.084	)	\times 10^{-9}	$	\\
7.80	--	7.85	&	2041    	&	$	(	9.838	\pm	0.216	+	0.727	-	0.933	)	\times 10^{-10}	$	\\
7.85	--	7.90	&	1586    	&	$	(	7.645	\pm	0.191	+	0.911	-	0.645	)	\times 10^{-10}	$	\\
7.90	--	7.95	&	1288    	&	$	(	6.208	\pm	0.172	+	0.445	-	0.592	)	\times 10^{-10}	$	\\
7.95	--	8.00	&	997     	&	$	(	4.806	\pm	0.151	+	0.416	-	0.371	)	\times 10^{-10}	$	\\
8.00	--	8.10	&	1469    	&	$	(	3.540	\pm	0.092	+	0.327	-	0.306	)	\times 10^{-10}	$	\\
8.10	--	8.20	&	956     	&	$	(	2.304	\pm	0.074	+	0.201	-	0.253	)	\times 10^{-10}	$	\\
8.20	--	8.30	&	501     	&	$	(	1.207	\pm	0.054	+	0.129	-	0.098	)	\times 10^{-10}	$	\\
8.30	--	8.40	&	307     	&	$	(	7.399	\pm	0.422	+	0.632	-	0.726	)	\times 10^{-11}	$	\\
8.40	--	8.50	&	201     	&	$	(	4.844	\pm	0.342	+	0.407	-	0.437	)	\times 10^{-11}	$	\\
8.50	--	8.60	&	93      	&	$	(	2.241	\pm	0.232	+	0.226	-	0.283	)	\times 10^{-11}	$	\\
8.60	--	8.70	&	61      	&	$	(	1.470	\pm	0.188	+	0.174	-	0.125	)	\times 10^{-11}	$	\\
8.70	--	8.80	&	39      	&	$	(	9.399	\pm	1.505	+	2.493	-	1.996	)	\times 10^{-12}	$	\\
8.80	--	8.90	&	22      	&	$	(	5.302	\pm	1.130	+	0.433	-	0.596	)	\times 10^{-12}	$	\\
8.90	--	9.00	&	19      	&	$	(	4.579	\pm	1.051	+	0.458	-	0.392	)	\times 10^{-12}	$	\\
9.00	--	9.10	&	5       	&	$	(	1.205	\pm	0.539	+	0.480	-	0.250	)	\times 10^{-12}	$	\\

 \end{tabular}
 \end{ruledtabular}
 \end{table*}

\providecommand{\noopsort}[1]{}\providecommand{\singleletter}[1]{#1}%

\end{document}